\documentclass[11pt,a4paper]{article}

\usepackage[utf8]{inputenc}
\usepackage{a4wide}
\usepackage{amsmath}
\usepackage{graphicx}
\usepackage{authblk}
\usepackage{color}

\begin{document}
\title{Efficient dynamical correction of the transition state theory rate estimate for a flat energy barrier}


\author[1,2,3]{Harri M\"okk\"onen}
\author[1,2,4]{Tapio Ala-Nissila}
\author[1,3]{Hannes J\'onsson}

\affil[1]{\footnotesize{Department of Physics, Aalto University School of Science, P.O. Box 1100, FIN-00076 Aalto, Espoo, Finland}}
\affil[2]{COMP CoE, Aalto University School of Science, P.O. Box 1100, FIN-00076 Aalto, Espoo, Finland}
\affil[3]{Science Institute of the University of Iceland, VR-III, 107 Reykjav{\'\i}k, Iceland  \footnote{hj$@$hi.is}}
\affil[4]{Department of Physics, Brown University, Providence RI 02912-1843, U.S.A.}

\maketitle

\begin{abstract}
The recrossing correction to the transition state theory estimate of a thermal rate can be difficult to calculate when the 
energy barrier is flat. This problem arises, for example, in polymer escape if the polymer is long enough to stretch 
between the initial and final state energy wells while the polymer beads undergo diffusive motion back and forth over the barrier.
We present an efficient method for evaluating the correction factor by constructing a sequence of hyperplanes 
starting at the transition state and calculating the probability that the system advances from one hyperplane to another
towards the product.
This is analogous to what is done in forward flux sampling except that there the hyperplane sequence starts at the initial state.  
The method is applied to the escape of polymers with up to 64 beads from a potential well. 
For high temperature, the results are compared with direct Langevin dynamics simulations as well as forward flux 
sampling and excellent agreement between the three rate estimates is found. 
The use of a sequence of hyperplanes in the evaluation of the recrossing correction speeds up the calculation by an
order of magnitude as compared with the traditional approach. 
As the temperature is lowered, the direct Langevin dynamics simulations as well as the forward flux simulations become 
computationally too demanding, 
while the harmonic transition state theory estimate corrected for recrossings can be calculated without significant 
increase in the computational effort. 
\end{abstract}

\pagebreak

%
%
\section{Introduction}

Transitions inducted by thermal fluctuations in atomic systems such as chemical reactions, diffusion events and 
configurational changes are often much less frequent than atomic vibrations.  In order to estimate the rate of
such rare events, a direct dynamical simulation requires much too long a simulation time while a statistical approach can be applied because of the separation of time scales.
The rate theory developed by Eyring and Polanyi~\cite{eyring_31} and, in a different form, by Pelzer and Wigner~\cite{wigner_35} 
for chemical reactions, can be applied in many different contexts.  We will refer to this as transition state theory (TST).
It provides a method for estimating the rate of rare events by performing a statistical average over the fast, oscillatory motion
and focuses on the probability of significant transitions.
A key concept there is the transition state, the region of configurational space representing a bottleneck for the transition 
and corresponding to a free energy barrier. 
The basic assumption of TST is that the transition state is only crossed once.  If the system
makes it to the transition state and is heading away from the initial state, it is assumed that the trajectory ends up in a product 
state for an extended period of time, before a possible back-reaction can occur. 
This approximation can then be checked and corrected by calculating short time trajectories
started at the transition state to obtain the so called recrossing (or 'dynamical') correction~\cite{keck_67}.
It turns out that TST gives an overestimate of the transition rate and the correction factor is $\kappa \le 1$.
If the transition state is variationally optimised~\cite{keck_67}, the correction factor is as close to unity as possible.
This two-step procedure for obtaining the value of the rate offers great advantage over the direct calculation of the rate from
trajectories starting at the initial state. It can take an impossibly long time to simulate even one such reactive trajectory, 
while the trajectories needed for the correction to the TST estimate are short since they start at the transition state. 
We will refer to this approach as the two step Wigner-Keck-Eyring (WKE) procedure~\cite{jonsson_11}.

The TST estimate of a transition rate requires, in general, the evaluation of the free energy of the system using some 
thermal sampling. But, a simpler approach is to apply a harmonic approximation where the rate is estimated by 
identifying the maximum energy along the minimum energy path of the transition. This point corresponds to a 
first order saddle point on the energy surface and gives the activation energy. The entropic prefactor is obtained
by evaluating the vibrational modes at the saddle point and at the initial state minimum.
This simplified version of TST is referred to as harmonic transition state theory (HTST) \cite{Wert1949,Vineyard1957}.

A recrossing of the transition state can occur for two different reasons.  One is the fluctuating force acting on the 
system due to the heat bath. If such a force is large enough and acts in the direction opposite to the velocity of the system 
soon after it has 
crossed the transition state, the system can go back through the transition state to the part of configuration space 
corresponding to the initial state.
The other reason for a recrossing of the transition state is related to the shape of the potential energy surface.
If the reaction path is curved near the transition state the system can enter a repulsive region that creates a force on the 
system that sends it back to the initial state. The more recrossings that occur, the smaller $\kappa$ becomes. 
The recrossings are particularly important if the energy along
the reaction path in the region near the transition state is relatively constant, i.e the energy barrier is flat.

A different approach to the estimation of thermal transition rates was developed by Kramers \cite{kramers_40} 
and later generalised to multidimensional systems by Langer \cite{langer_69}.
There, a harmonic approximation to the energy surface is used and
a statistical estimate is made for the recrossings due to the fluctuating force from the heat bath.
The advantage of this approach is that some of the recrossings are taken into account in the rate estimate without 
requiring a dynamical recrossing correction. 
A disadvantage as compared to the two step WKE procedure is that some of the recrossings are not included, in particular not the
ones resulting from the shape of the potential energy surface.
Another disadvantage of the Kramers/Langer approach is a harmonic approximation of the energy surface
in the direction of the reaction path at the transition state. As a result, the rate is estimated to vanish 
if the energy barrier is flat, i.e. if the second derivative of the energy along the reaction path is zero.
Such flat top energy barriers can occur in various applications. 

One example of a flat barrier problem is the transition of a polymer from a potential well, the so-called polymer escape problem \cite{Park1999,Sebastian2000,Sebastian2000a, Sebastian2006,Sebastian2010, Lee2001, Lee2001a, Paul2005}.
Experimental examples of systems of this sort include polymer translocation \cite{Palyulin2014, Muthukumar2011}, 
where a polymer is crossing a membrane through a pore \cite{Kasianowicz1996}, or narrow  $\mu$m-scale channels with traps \cite{Han1999}. Recent experiments by Liu \textit{et al.} involve the escape of a DNA molecule from an entropic cage \cite{Liu2014}. Similar translocation and escape processes are common in cell biology and have possible bioengineering applications, such as DNA sequencing \cite{Howorka2001} and biopolymer filtration \cite{Mikkelsen2011}. 

In a recent study of a model polymer escape problem, the application of HTST followed by dynamical corrections was shown 
to give accurate results as compared to direct dynamical simulations, while the Kramers-Langer approach gave a significant
underestimate of the transition rate for the longer polymers \cite{Mokkonen2014,Mokkonen2015}.  
Flat energy barriers are also common in magnetic transitions involving the temporary domain wall mechanism \cite{bessarab_13,bessarab_14}. 

The evaluation of the recrossing correction in the WKE procedure can involve large computational effort 
for extended, flat energy barriers.  This occurs, for example, for the escape of long polymers that are
long enough to stretch between the initial and final state energy wells while the polymer beads undergo diffusive motion back and forth over the barrier.
Well established procedures exist for the evaluation of the recrossing correction from trajectories that start and the 
transition state and eventually make it to either the final state or back to the initial state (see, for example, 
Ref. \ref{Voter1985}).
However, the diffusive motion along the
flat energy barrier can make such trajectories long and the calculation computationally demanding.

We present here an approach for calculating the recrossing correction in such challenging problems by using 
a procedure that is similar to so-called forward flux sampling (FFS) \cite{Allen2006,Allen2006a,Allen2009}
where a sequence of hyperplanes is constructed
and trajectories are calculated to estimate the probability that the system advances from one hyperplane to the next.
While the FFS method has been proposed as a way to calculate transition rates starting from the initial state and ending 
at a final state, we start the hyperplane sequence at the transition state and evaluate the probability that the system 
makes it all the way to the final state given that it starts at the transition state. 
This turns out to be a more efficient procedure than the standard method for evaluating 
the recrossing correction by calculating individual trajectories
that make it all the way from the transition state to the final state \cite{Voter1985}.
We present results on the computational efficiency of these different methods for estimating the escape rates of polymers 
with up to 64 beads, where there is a pronounced flat barrier.

The article is organised as follows: In the following section, Sec. \ref{sec:methods}, the test problem, parameters and numerical methods are described. The results 
of the various calculations are presented and the efficiency compared in Sec. \ref{sec:results}. The article concludes with
a summary and discussion in Sec. \ref{sec:discussion}. 

%
%
\section{System and methods used} \label{sec:methods}

\subsection{Description of the system}

\begin{figure}
\centering
\includegraphics[width=120mm]{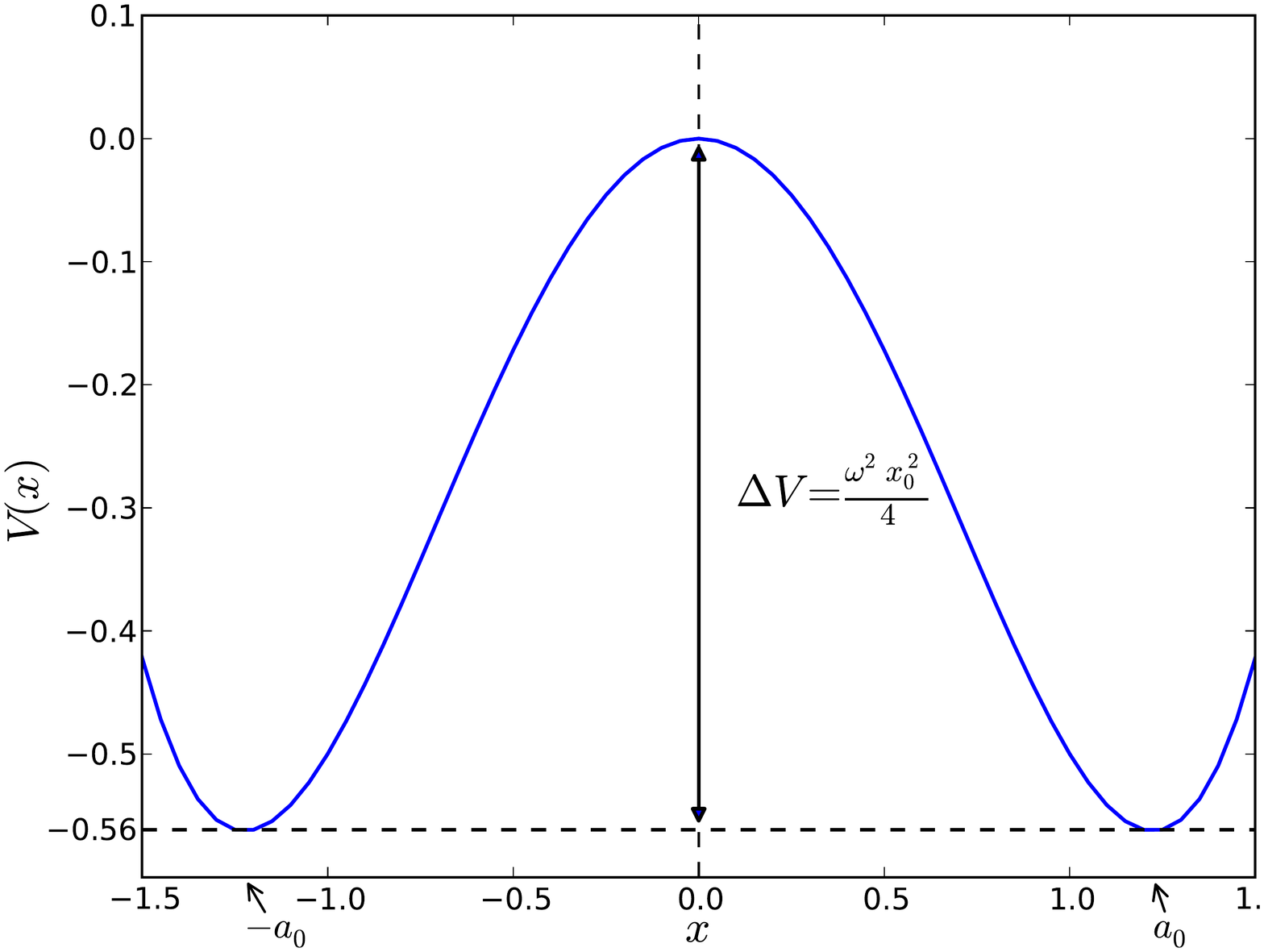}
\caption{
The external potential of Eq. \eqref{eq:externalpotential}. The maximum of height $\Delta V = \omega^2 a_0^2 / 4 \approx 0.56$ is located at $x = 0$ and the symmetric minima are located at $x = \pm a_0 \approx \pm 1.22$. 
The  initial state, I, is confined to the left well $x < 0$ and the right well $x > 0$ corresponds to the final state, F. 
} 
\label{fig:externalpotential}
\end{figure}

The polymer is described by a set of $N$ identical beads that are connected to two neighboring beads except for the end points.
The configuration of the polymer is described by the coordinates of the beads $\mathbf{r} :=  \{ r_n \}_{n=1}^N$.
The centre of mass of the polymer is $X_0 = \frac{1}{N} \sum_{n=1}^N r_n$.
The dynamics of the beads is described by the Langevin equation where for the $n$th bead at time $t$ 
\begin{equation} 
m \ddot{r}_n(t) +  \gamma \dot r _n(t)  + \nabla_n [V(r_n(t)) + U] = \sqrt{2 \gamma k_B T } \xi_n (t), \label{eq:langevin}
\end{equation} 
where $m$ is the mass of a bead, $\gamma$ the friction coefficient, $V(r_n)$ the external potential, $U$ the interaction 
potential between adjacent beads, $k_BT$ the thermal energy, and $\xi_n (t)$ a Gaussian random force satisfying $\langle 
\xi_n (t) \rangle = 0$ and $\langle \xi_n (t) \xi_m (t') \rangle = \delta (t - t') \delta_{n,m}$. 
The interaction between adjacent beads is given by a harmonic potential function 
\begin{equation} 
U =  \sum_{n=1}^{N-1}(K/2)(r_n - r_{n+1})^2 .
\end{equation}
The resulting contribution to the force on bead $n$ is
\begin{equation} 
- \nabla_n U = - K (r_{n-1} + r_{n+1} - 2r_n).
\end{equation}
The external potential, $V(x)$, shown in Fig. \ref{fig:externalpotential}, is a quartic double well
\begin{equation} 
V(x) = - \frac{\omega^2}{2} x^2 + \frac{\omega^2}{4 a_0^2} x^4,\label{eq:externalpotential}
\end{equation}
where $\pm a_0$ gives the locations of the minima. The energy has a maximum at $x=0$ where
the curvature of the potential energy function is $\omega^2$. This is the same external potential that was used in Refs. \ref{Mokkonen2015} and \ref{Lee2001}. The total potential energy of the system is $ \sum_{n=1}^{N}V(r_n) + U$.

\subsection{Direct simulations}

From direct Langevin dynamics (LD) simulations starting at the equilibrated initial state, the escape probability 
can be evaluated by observing the time it takes for the system to reach the final state in a number of statistically
independent trajectories.  A trajectory is taken to have reached the final state when the centre of mass 
is half way between the maximum and the final state minimum, $X_0 > a_0/2$.
If $t_i$ is the time of the escape event occurring in trajectory $i$, then the
thermally averaged probability that a transition has occurred after time $t$ is
\begin{equation}
P_\mathrm{esc} (t) = (1/N_\mathrm{traj}) \sum_{i=1}^{N_\mathrm{traj}} \theta(t - t_i), \label{eq:pesc}
\end{equation} 
where $N_\mathrm{traj}$ is the number of trajectories simulated, and $\theta(t -t_i)$ the Heaviside step function. 
The escape rate is then given by 
\begin{equation}
\mathcal R_\mathrm{LD} = \left. \frac{d P_\mathrm{esc} (t)}{dt} \right., \label{eq:dynamicalrate}
\end{equation} 
where the derivative is determined by fitting to the linear region in the function $P_\mathrm{esc} (t)$.


\subsection{Forward flux sampling}

Forward flux sampling is a class of methods based on a series of hyperplanes, $\lambda_0, \lambda_1, \dots, 
\lambda_n$ placed between the initial and final states \cite{Allen2006, Allen2006a, Allen2009}. 
The rate constant is calculated by sampling the dynamics between the hyperplanes. 
We will give a brief review of the method here. 
For a more detailed description of the method the reader is referred to the review article by Allen \textit{et al.} \cite{Allen2009}.

The rate constant is obtained in FFS as \cite{Allen2006}
\begin{equation}
\mathcal R_\mathrm{FFS} =  \frac{\bar \Phi_{I,0}}{\bar h_I} P(\lambda_n, \lambda_0),
\end{equation}
where $\bar \Phi_{I,0}/\bar h_I$ is the initial flux across the first plane $\lambda_0$ towards the final state, 
and $P(\lambda_n | \lambda_0)$ is the probability that the system reaches plane $\lambda_n$ given it was initially at 
$\lambda_0$. The initial flux is calculated by simulating a long trajectory in the initial state for a time $t_\mathrm{init}$ and counting the number of crossings $q$ of the first hyperplane, $\lambda_0$, with the normal component of the 
velocity pointing towards the final state. Therefore, the initial flux is $\bar \Phi_{I,0}/\bar h_I = q / t_\mathrm{init}$. 

The configuration at each of the crossing events of the $\lambda_0$ hyperplane serves as an initial configuration for a new trajectory which is run until the next interface $\lambda_1$ is reached, or the trajectory returns to the initial state by crossing 
$\lambda_0$. The probability $P(\lambda_1 | \lambda_0)$ is estimated as the fraction between the number of successful trajectories and the number of all trajectories initiated from $\lambda_0$. 
The final configurations of the successful trajectories are used as initial points to sample the probability $P(\lambda_2 | \lambda_1)$, which is then a ratio of the trajectories that reach $\lambda_2$ to those that return to the initial state by crossing $\lambda_0$. The procedure is repeated until all the hyperplanes have been sampled and the probability 
\begin{equation}
P(\lambda_n | \lambda_0) = \prod_{i=0}^{n-1} P(\lambda_{i+1} | \lambda_i) \label{eq:ffsprob}
\end{equation}
can be computed.
 

\subsection{HTST and recrossing corrections}

The evaluation of the HTST estimate of the escape rate requires finding the first order saddle point on the energy surface 
defining the transition state and evaluating the vibrational frequencies from eigenvalues of the Hessian at the saddle point and 
the initial state minimum. In order to find the saddle point, 
the nudged elastic band (NEB) method \cite{Jonsson1998,Henkelman2000,Henkelman2000a}
was used to determine the minimum energy path (MEP) for the transition. The point of maximum energy along the MEP
is the relevant saddle point. 

The Hessian matrix was evaluated at the minimum and at the saddle point using finite differences of the 
forces on the beads, and the eigenvalues calculated.
The HTST estimate of the transition rate is \cite{Wert1949,Vineyard1957}
\begin{equation}
\mathcal{R}_\text{HTST} = \frac{1}{2 \pi \sqrt{\mu_\perp}} \sqrt{\frac{\prod_{i=1}^N \lambda^0_i}{\prod_{i=2}^{N} \lambda^\ddagger_i} }e^{-\Delta E / k_B T},\label{eq:ratehtst}
\end{equation}
where $\mu_\perp$ is the reduced mass,
and $\lambda^0_i$ and $\lambda^\ddagger_i$ are the eigenvalues of the Hessian matrices at the minimum and at the saddle point, respectively. The activation energy, $\Delta E$, is the 
potential energy difference between the minimum and the saddle point. 
The negative eigenmode at the first order saddle point is labeled as $i=1$ and is omitted from the product in the denominator. 
In HTST the transition state is chosen to be the hyperplane containing the first order saddle point and having a normal
pointing in the direction of the unstable mode, i.e. the eigenvector corresponding to the negative eigenvalue.


The recrossing correction can be estimated by starting trajectories at the transition state and observing  
recrossings of the transition state until the trajectory ends up in either the initial or final state.
Voter and Doll \cite{Voter1985} have described a method for computing the correction factor, $\kappa$, 
from an ensemble of such trajectories.
The corrected transition rate is $\mathcal{R}_\text{HTST+VDDC} = \kappa \mathcal{R}_\text{HTST}$
where VDDC stands for dynamical correction evaluated following the procedure of Voter and Doll. 

The escape rate of polymers has previously been studied \cite{Mokkonen2015} using HTST followed by 
recrossing corrections evaluated using the method of Voter and Doll \cite{Voter1985}.


\subsection{Hyperplane sequence for recrossing correction}

We propose here an efficient method for calculating the recrossing correction using a sequence of hyperplanes, 
analogous to the formulation of the FFS method. Here, however, the initial hyperplane is placed at the transition state.
The sequence of parallel hyperplanes then leads to a final hyperplane near the minimum on the energy surface
corresponding to the final state. Instead of sampling the initial flux, we equilibrate the system within the transition state
hyperplane to generate uncorrelated samples and afterwards assign a random velocity from the Maxwellian distribution 
$P(v_i) \propto \exp (-m v_i^2 / 2 k_B T)$ to each degree of freedom. 
If the net velocity $\sum_{i=1}^N v_i$ of the system is negative, the velocities are reversed $v_i \rightarrow -v_i$ to describe a trajectory heading towards the final state. 
The dynamical correction factor is computed according to Eq. \eqref{eq:ffsprob} as
\begin{equation}
\kappa = P(\lambda_n | \lambda_0) = \prod_{i=0}^{n-1} P(\lambda_{i+1} | \lambda_i), \label{eq:ffsdc}
\end{equation}
where the initial hyperplane $\lambda_0$ is located at the transition state and 
the final hyperplane $\lambda_n$ is near the final state minimum.
We will refer to this as FFDC method for calculating the recrossing correction.

As illustrated below, the use of a hyperplane sequence and short time trajectories between the hyperplanes can reduce the 
computational effort involved in determining the recrossing correction factor as compared with the use of long trajectories that
make it all the way from the transition state to the final state.     


\subsection{Parameters and numerical methods} 

The Langevin trajectories of Eq. \eqref{eq:langevin} were calculated using the Br\"unger-Brooks-Karplus integration scheme \cite{Brunger1984} with a time step of $\Delta t = 0.005$. The number of beads $N$ in the polymer chains
was in the range $N \in \{1, \dots, 64\}$. The parameters were chosen to be $\gamma = 1.0$ and $m=1.0$.
The parameters for the external potential of Eq. \eqref{eq:externalpotential} were chosen to be 
$\omega^2 = 1.5$ and $a_0^2 = 1.5$. 
The same values of the parameters were used in Refs. \ref{Lee2001} and \ref{Mokkonen2015}. 

If the units of length, mass and energy are chosen to be $l_0 = 1.02$ nm, $m_0 = 1870$ amu, corresponding to a double stranded DNA, and 
the unit of energy is $k_B T$ at $T = 300$ K, the unit of time becomes $t_0 = \sqrt{m_0 l_0^2 / k_B T}  =27.9$ ps. 
In the direct dynamical simulations to determine the escape rate a total of 1 000 to 240 000 trajectories 
were used depending on the chain length. 

In the calculations of the MEP using the NEB method
several images of the polymer were placed between the initial and final states and connected with harmonic springs. 
The energy was then minimised using the projected velocity Verlet integration \cite{Jonsson1998}. 
The precise location of the saddle point was found by minimising the force acting on the highest energy image using the Newton-Rhapson method. The spring constant used in the NEB calculations was $k_\mathrm{NEB} = 8.2$. The number of images, $P$, was typically chosen to be between 9 and 19, but for larger values of $N$, $P = N/2$ was sometimes used.

In the FFS method, the number of hyperplanes between the initial and final states was $n=10$.
The initial hyperplane was placed at $\lambda_0 = -a_0$ and the final hyperplane $\lambda_n = a_0/2$. 
For the longer chains, additional hyperplanes were used, up to 20. To obtain the escape rate, 100 000 initial points were sampled and the number of trajectories started from each one. To obtain the statistical error in the escape rate for efficiency analysis, the FFS calculation was repeated with 10 000 trajectories started from each hyperplane and the standard deviation 
of the result computed. The statistical error estimate (standard error of the mean) was then evaluated as 
the standard deviation divided by $\sqrt{N_s}$, where $N_s$ is the number of samples. 
This was repeated until a similar level of accuracy was reached as for other methods for estimating the escape rate. 

In the FFDC calculations of the recrossing factor using a hyperplane sequence,
10 000 initial configurations were generated by equilibrating the system within the transition state.
Subsequently, 10 000 trajectories were generated from each plane to obtain the probability in Eq. \eqref{eq:ffsdc}. 
The same error estimate (standard error of the mean) was used as with the FFS calculation. 
The statistical error in $\kappa$ was computed by repeated runs until the desired level of accuracy was reached.  

%
%
\section{Results} \label{sec:results}

The escape rate was calculated at two different temperature values. At the higher temperature, $T=1.0$, the
escape rate is high enough that direct dynamical simulations are possible. This provides a good test 
for the accuracy of the various methods.  The lower temperature,  $T=0.5$, is more representative of a 
practical situation where the escape rate is so low that a direct dynamical simulation is not practical.
The FFS method also turns out to require excessive computational effort in that case, much more than
the HTST+FFDC approach.

\subsection{Escape rate at $T=1.0$}

\begin{figure}[!t]
\centering
\includegraphics[width=120mm]{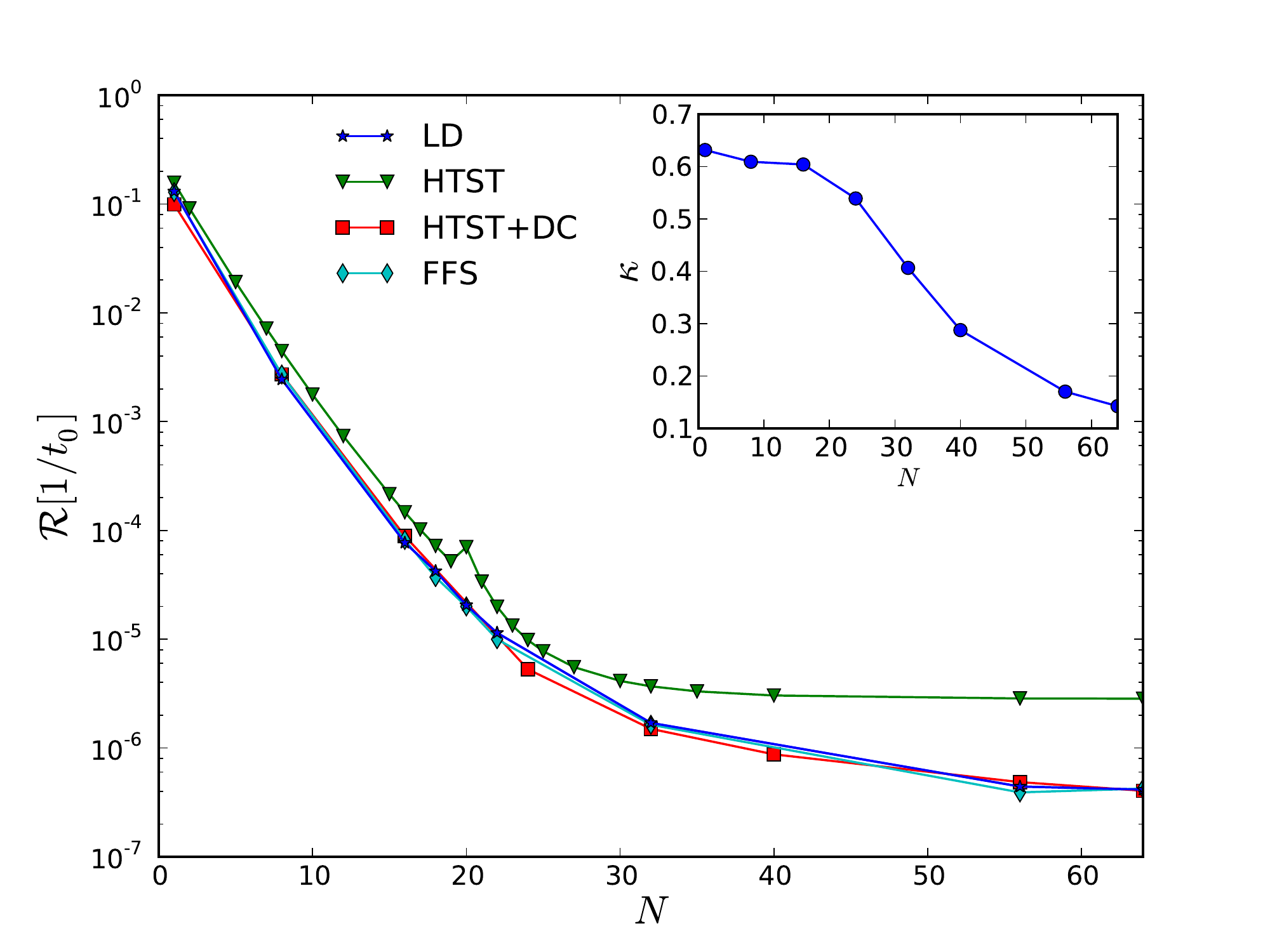}
\caption{
Escape rate for polymers obtained by direct Langevin dynamics simulations (stars), forward flux sampling (diamonds),
HTST (triangles), and HTST followed by recrossing correction HTST+FFDC (squares). The inset shows the correction factor 
$\kappa$ of Eq. \eqref{eq:ffsdc}.
} 
\label{fig:rates_htst}
\end{figure}

The escape rate of the polymer was computed for a temperature of $T=1.0$, a relatively high temperature,
with three different methods: 
direct Langevin dynamics (DLD) simulation, the forward flux sampling method, and with HTST followed by 
a recrossing correction.  Quantitive agreement between the results obtained by the various methods 
was obtained as shown in Fig. \ref{fig:rates_htst}. 

%

\begin{figure}[!t]
\centering
\includegraphics[width=120mm]{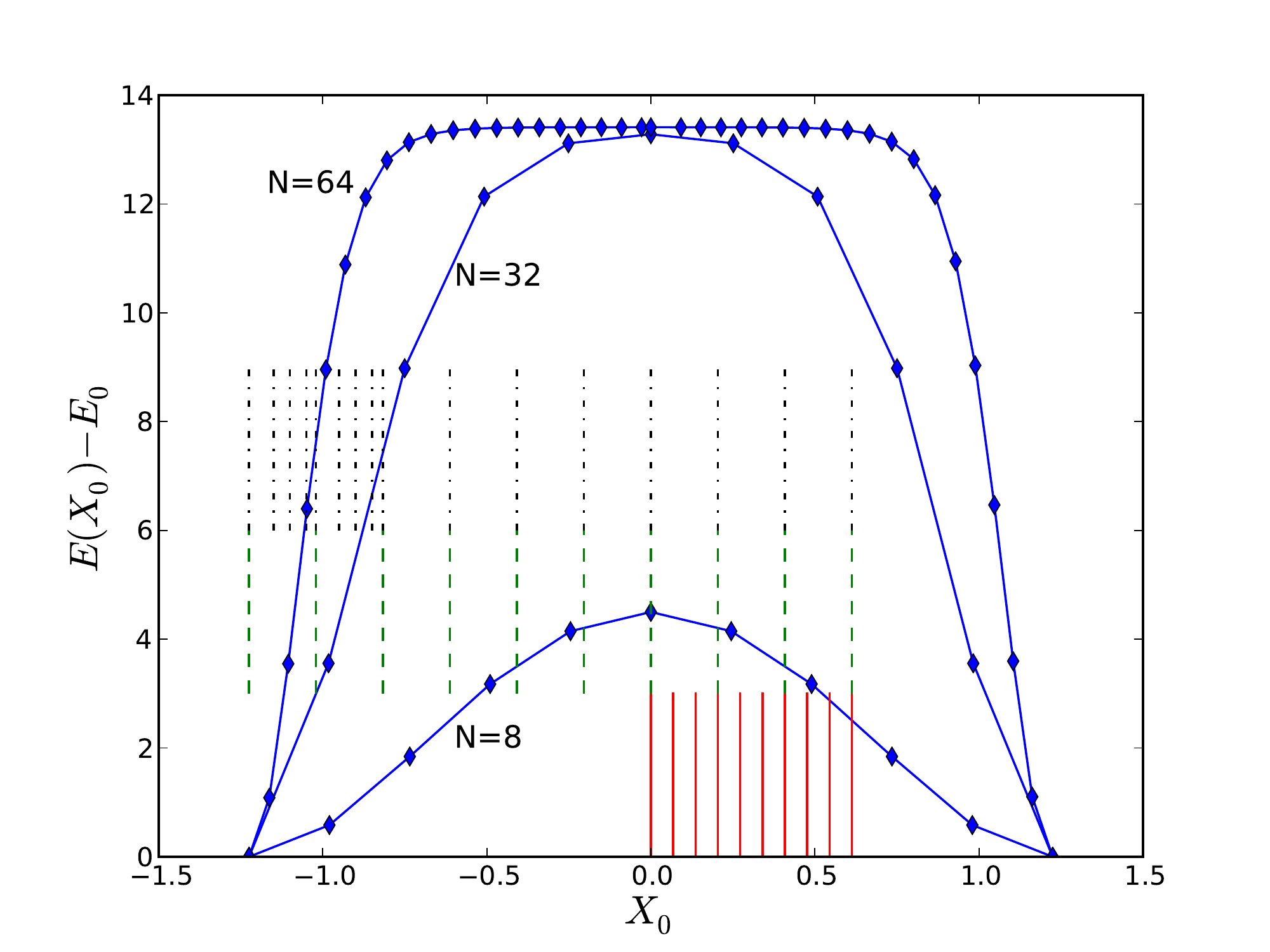}
\caption{
Energy along the minimum energy path as a function of the centre of mass for chains with $N\ =\ 8,\ 32$ and
$64$ beads. For ca. $N=32$ the height of the energy barrier reaches a plateau and its top becomes flat. The solid red lines represent the positions of the hyperplanes used in the FFDC calculation of $\kappa$, Eq. \eqref{eq:ffsdc}. The dashed green lines represent the hyperplanes used in the FFS calculation of the escape rate when the 
hyperplanes are positioned with equal spacing. 
The dotted and dashed black lines present the optimised hyperplane positions for FFS for the chain length $N= 64$. 
The hyperplane positions are optimised in such a way that six extra hyperplanes are added to the region where the 
slope of the energy barrier is steep. 
} \label{fig:barriers}
\end{figure}

The HTST rate shows a peak at $N=20$ which is due to the smallest positive eigenvalue at the saddle point approaching zero and causing divergence in the rate estimate, Eq. \eqref{eq:ratehtst}. 
Anharmonic corrections \cite{Mokkonen2015,Lee2001} are computed for this mode but they do not completely remove the peak. The HTST estimate of the rate saturates to a constant value in the region $N > 32$. 
This is because the height of the effective energy barrier, the maximum along the MEP, 
saturates as the barrier starts to flatten out as shown in Fig. \ref{fig:barriers}. In this region the influence of the recrossing
correction becomes particularly relevant. 
The hyperplanes used in the FFS calculations, and in the FFDC calculations (Eq. \eqref{eq:ffsdc}), are also shown. 
To optimise the performance of the FFS method, extra planes were added to the region where the slope of the energy barrier 
is steep. 

Comparison of the efficiency of the three methods applied to a system at temperature of $T=1.0$
for polymers of lengths $N=8$ and $N=64$ is shown in Table \ref{tbl:efficiency}. 
The computational cost for a desired level of accuracy was measured by counting the number of energy and force evaluations. 

\begin{table}[!t]
\centering
  \begin{tabular}{|c|c|c|}
  \hline
   Method &  $\Delta \mathcal R / \mathcal R$ & \# func. eval. \\
  \hline
  \hline
  \multicolumn{3}{|c|}{$N=8$} \\ 
  \hline
  Direct LD           & 6 \%   &$ 6.4 \times 10^{9} $ \\
  FFS (10 planes)& 6 \%   & $ 5.0 \times 10^{8} $ \ \\
  HTST+VDDC    &  3 \%  & $ 8.0 \times 10^{7}$ \\  
  HTST+FFDC     &  4 \%  & $8.0 \times 10^{6}$ \\  
  \hline
  \multicolumn{3}{|c|}{$N=64$} \\ 
  \hline
  Direct LD            & 10 \% & $ 3.0  \times 10^{13} $ \\
  FFS (10 planes) & 17  \% & $  9.0 \times 10^{11} $ \ \\
  FFS (16 planes) &  10  \% & $ 7.0  \times 10^{11} $ \ \\
  HTST+VDDC     &  9 \%  & $ 2.6 \times 10^{11}$ \\  
  HTST+FFDC      & 3 \%  & $  2.88 \times 10^{10}$ \\  
  \hline
  \end{tabular}
  \caption{
Relative errors $\Delta \mathcal R / \mathcal R$ of the escape rate and the number of the energy/force evaluations required for each method for polymers of length $N=8$ and $N=64$ at a temperature of $T=1.0$. 
The HTST+FFDC converges to an small relative error with the least computational effort, about an order of 
magnitude less than HTST+VDDC which in turn is about an order of magnitude more efficient than FFS.
At this relatively high temperature, the direct Langevin dynamics simulation can be carried out to obtain an estimate
of the escape rate and it turns out to be one to two orders of magnitude less efficient than FFS, depending on the length 
of the polymer. 
} \label{tbl:efficiency}
\end{table}

For polymers of length $N=8$, the FFS is an order of magnitude faster than direct Langevin dynamics. 
The HTST with recrossing correction computed using the method of Voter and Doll \cite{Voter1985}, HTST+VDDC, 
is in turn an order of magnitude faster than FFS.  
HTST with recrossing correction computed using the hyperplane sequence of Eq. \eqref{eq:ffsdc} is, furthermore, an
order of magnitude faster than HTST+VDDC. 
For polymers of length $N=64$, the ratios of efficiency are similar to the $N=8$ chain, except the direct Langevin dynamics
simulation is two orders of magnitude slower than FFS, and the efficiency of HTST+VDDC is closer to the efficiency of FFS. 

The efficiency of FFS can be optimised by adjusting the number and location of the hyperplanes 
and adjusting the number of trial runs for each plane \cite{Allen2009,Borrero2008}. 
Additional hyperplanes were introduced to decrease the spacing between them where the forward flux 
$P(\lambda_{i+1} | \lambda_i)$ turned out to be small. For polymers of length $N=8$ and $N=32$, the addition of extra planes to the small flux region did not improve the computational efficiency. 
For polymers of length $N=64$, six additional planes were added to the region where the potential gradient is steep 
(see Fig. \ref{fig:barriers}) and  the forward flux $P(\lambda_{i+1} | \lambda_i)$ is small. Table \ref{tbl:efficiency} shows that 
for $N=64$, the optimised FFS method produces a smaller relative error than the unoptimised one, with smaller number of energy/force evaluations.

\subsection{Escape rate at $T=0.5$}

At the lower temperature, which is more representative of a typical situation, the direct dynamical simulation for polymers
with $N=64$ becomes computationally too demanding.
Also, the use of the FFS method becomes difficult at this temperature since most of the FFS simulations fail. 
At some point none of the trajectories make it to the next plane, and the simulation comes to a stop. 
In order to improve the performances of the FFS method, 10 additional hyperplanes were added to the small flux region 
(where the slope of the energy barrier is steep), but still the method failed most of the time. 
The error estimate reported in Table \ref{tbl:halftemp} for FFS is obtained using only successful FFS samples, 
so it represents a lower bound for the number of energy/force evaluations 
needed to obtain an estimate with a 20\% error estimate. 
Since the standard error of the mean scales roughly as $\sim 1/ \sqrt{N_s}$ with the number of samples, $N_s$, 
this value can be extrapolated to estimate the number of energy/force evaluations required to 
obtain an estimate with a relative error of 3\%. This gives an estimate of $3.2 \times 10^{12}$. 
The HTST+FFDC method about two orders of magnitude more efficient than FFS in this case.

\begin{table}[!t]
\centering
  \begin{tabular}{|c|c|c|}
  \hline
   Method &  $\Delta \mathcal R / \mathcal R$ & \# func. eval. \\
  \hline
  FFS (20 planes)& 20 \% & $ 9.3 \times 10^{11} $  \\
  HTST+FFDC &  3 \%  & $ 6.0 \times 10^{10}$ \\  
  \hline
  \end{tabular}
  \caption{Comparison of computational efficiency of FFS and HTST+FFDC at $T=0.5$ for polymers with $N=64$. 
Even with addition of extra hyperplanes in the small flux region, most of the FFS simulations fail in that at some point
none of the trajectories make it to the next plane.
The number of energy/force evaluations reported here includes only the successful FFS calculations, 
so it present a lower bound for the number of function evaluations needed for the this level of accuracy.} 
\label{tbl:halftemp}
\end{table}


%
%

\section{Summary and discussion} \label{sec:discussion}

In this article, we have proposed an efficient method for evaluating the recrossing correction to transition state theory
which is particularly useful for systems with a flat energy barrier, i.e. where the energy along the reaction path
is nearly constant at the transition state.
The method is benchmarked in calculations of the
escape rate of polymers with up to 64 beads to provide recrossing corrections to harmonic transition state 
theory estimate of the rate. 
At high temperature, the results are compared with results using
using direct Langevin dynamics simulations, as well as forward flux sampling, and harmonic transition state theory with 
recrossing corrections evaluated in a traditional way.
The computational efficiency of these various methods was compared by counting the number evaluations of the energy and
force needed to reach a desired level of accuracy in the rate estimate. 
The method is shown to be accurate and significantly more efficient than the other methods.
This is even more so at a lower temperature which represents a more typical situation.

The efficiency of the HTST+FFDC methods stems from the fact that transition state theory is used to identify the 
bottleneck for the transition and the time scale problem of simulating a long trajectory starting at the initial state and
eventually making it over to the final state is avoided.  The rate estimate obtained this way is approximate, though, 
because of the no recrossing approximation of transition state theory. By carrying out calculations of short time
trajectories starting at the transition state a correction for this approximation can be made. Since the trajectories are going
downhill in energy, they are relatively short. The forward flux method, however,
is computationally more demanding because it relies on trajectories that go uphill in energy, and only a small fraction of
the trial trajectories do so. Furthermore, a key issue is the orientation of the hyperplanes which is not specified in the 
forward flux methodology. If the orientation of the hyperplane representing the bottleneck 
of the transition is not right, the sampling will be problematic and the rate estimate likely incorrect. The variational
principle of transition state theory \cite{keck_67} can, however, be used to find the optimal orientation of the
hyperplane \cite{johannesson_01} which in turn identifies the optimal mechanism of the transition.



\section*{Acknowledgements}

This work was supported by the Academy of Finland through the FiDiPro program (H. J. and H.M., grant no. 263294) and the COMP CoE (T. A-N, grants no. 251748 and 284621). We acknowledge computational resources provided by the Aalto Science-IT project and CSC -- IT Center for Science Ltd in Espoo, Finland.

\end{document}